\documentclass[12pt]{iopart}
\usepackage{graphicx}
\usepackage{cite}
\usepackage{amsfonts}
\begin{document}

\title[Response function of the disordered Ising model]
      {Dynamical response function of the disordered kinetic Ising model}
\author{Haye Hinrichsen}
\address{Universit\"at W\"urzburg\\
	 Fakult\"at f\"ur Physik und Astronomie\\
         D-97074 W\"urzburg, Germany}
\ead{hinrichsen@physik.uni-wuerzburg.de}

\begin{abstract}
Recently Baumann \textit{et al.} [arXiv:0709.3228v1] studied the phase-ordering kinetics of the two-dimensional Ising model for $T<T_c$ with uniform spatially quenched disorder by Monte-Carlo simulations. They found that the two-time response and correlation functions are in agreement with the predictions of local scale invariance generalised to $z\neq 2$. The present paper shows why this is actually not true and suggests an alternative approach which leads to a much better agreement with the numerical results.
\end{abstract}

\submitto{Journal of Statistical Mechanics: Theory and Experiment}

% Explanation of PACS numbers:
% 05.50.+q: Lattice theory and statistics
% 05.70.Ln: Nonequilibrium and irreversible processes
% 64.60.Ht: dynamic critical phenomena

\parskip 2mm 

%==========================================================================
\section{Introduction}
%==========================================================================

A ferromagnetic systems starting with a disordered initial state exhibits phase ordering kinetics, forming ordered domains whose linear size grows with time as $\xi \sim t^{1/z}$, where $z$ is the dynamical exponent. For homogeneous systems with a non-conserved order parameter one has $z=2$~\cite{Bray94}, while in systems with spatially quenched disorder the dynamical exponent is found to vary continuously with the intensity of the disorder. The probably simplest example of such a system is the two-dimensional ferromagnetic Ising model with quenched disorder~\cite{PaulEtAl04}. It is defined by the Hamiltonian
\begin{equation}
\mathcal{H}\;=\;-\sum_{(i,j)}J_{ij}\sigma_i\sigma_j\,,\qquad \sigma_i=\pm 1\,,
\end{equation}
where the sum runs over all nearest neighbours of a two-dimensional square lattice with periodic boundary conditions. The coupling constants $J_{ij}=J_{ji}$ are uncorrelated random variables distributed uniformly over the interval
\begin{equation}
J_{ij} \in [1-\epsilon/2\,,\,1+\epsilon/2]\,,
\end{equation}
where $\epsilon\in[0,2]$ is a parameter controlling the strength of the disorder. The model evolves by conventional heat bath dynamics, i.e., for each local update a site $i$ is randomly selected and the spin $\sigma_i$ is oriented according to the local field $h_i=\sum_{(i,j)}J_{ij}\sigma_j$ of the neighboring spins by setting
\begin{equation}
\sigma_i \;:=\; 
\left\{
\begin{array}{ll}
+1 & \mbox{ with probability } \frac{e^{h_i/kT}}{e^{h_i/kT}+e^{-h_i/kT}}\\
-1 & \mbox{ otherwise. }                 
\end{array}
\right.
\end{equation}
The model exhibits a second-order phase transition between a paramagnetic and a ferromagnetic phase at some critical temperature $T_c(\epsilon)>0$ which depends on the intensity of the disorder. In the ferromagnetic phase $T<T_c(\epsilon)$, starting with a random initial configuration, one observed phase ordering with a dynamical exponent $z>2$. According to phenomenological scaling arguments~\cite{PaulEtAl04} and field-theoretic studies~\cite{SchehrLeDoussal05,SchehrRieger05} its value is given by\footnote{Note that this relation is still debated in the literature. In particular, Huse and Henley~\cite{HuseHenley85} conjectured that domains grow as a power of $\log t$. However, the simulations performed in the present work seem to be consistent with Eq.~(\ref{ZFormula}).}
\begin{equation}
\label{ZFormula}
z \;=\; 2+\frac{\epsilon}{T}.
\end{equation}
For example, for $T=1$, -- a temperature well below $T_c$ that will be considered throughout this paper -- the dynamical exponent varies between 2 and 4 depending on the strength of the disorder $\epsilon$. 

Phase ordering in disordered ferromagnets gives rise to \textit{physical ageing}, charac\-terised by a non-exponential relaxation, dynamical scaling, and breaking of time-translational invariance~\cite{CalabreseGambassi05}. An interesting quantity, which allows one to study ageing phenomena, is the \textit{response function}
\begin{equation}
R(t,s;{\vec r}) \;=\; 
\left.\frac{\delta\langle\phi(t,{\vec r})\rangle}{\delta h(s,0)}\right|_{h\to 0}\,,
\end{equation}
which describes the averaged spatio-temporal response of the magnetisation $\phi$ to a \textit{local} perturbation by a weak magnetic field $h$. The response function can be measured by applying a weak field at a particular site at time $s$ and monitoring subsequent spin flips caused by the field at distance ${\vec r}$ at some later time $t>s$. Since in disordered ferromagnetic Ising model with heat bath dynamics a positive external field $h(s,0)>0$ can only turn spins up but never down, the response function is strictly positive in the present case. 

Ageing phenomena are characterised by dynamical scaling. This postulate implies that the response function obeys the scaling form
\begin{equation}
R(t,s;{\vec r}) \;=\; 
s^{-1-a} \, \mathcal{R} \left(\frac{t}{s},\, \frac{{\vec r}}{(t-s)^{1/z}}\right)\,,
\end{equation}
where $\mathcal{R}$ is a scaling function and
\begin{equation}
a=1/z.
\end{equation}
During the past decade there have been various attempts to generalise the concept of dynamical and anisotropic scaling to some kind of local scale invariance~\cite{Henkel02}, in a similar sense as conformal invariance generalises global scale invariance of two-dimensional equilibrium phase transitions to local scale transformations. Such theories of local scale invariance (LSI) together with explicit representations of its generators make non-trivial predictions about the specific form of scaling functions. Recently, LSI was developed further as to include systems with $z \neq 2$~\cite{HenkelBaumann07b}. As a first application to a non-linear model with $z \neq 2$, Baumann et al.~\cite{BaumannEtAl07} applied this generalised version of LSI to the disordered ferromagnetic Ising model, finding the predictions in good agreement with the numerical results. In contrast to these results, the aim of the present work is

\begin{enumerate}
\item[(a)]  to show that the predictions of generalised LSI, as quoted in~\cite{BaumannEtAl07}, are in principle incompatible with the actual response function of the disordered Ising model,
\item[(b)]  to demonstrate this incompatibility numerically by a \textit{direct} measurement of the response function,
\item[(c)]  to suggest an alternative approach with algebraically distributed waiting times which yields consistent results in much better agreement with the numerical data. 
\end{enumerate}

\noindent
The present results for $T<T_c$ complement previous studies Pleimling, Gambassi, Corberi, Lippiello and Zannetti~\cite{PleimlingGambassi05,CorberiEtAl07} who have questioned the applicability of LSI to the critical Ising model without quenched disorder at criticality. A similar discussion concerning the application of LSI to critical directed percolation without quenched disorder can be found in Refs.~\cite{Hinrichsen06,BaumannGambassi07}. 

%==========================================================================
\section{Test of the predictions of LSI}
%==========================================================================

According to Henkel {\it et al.}~\cite{HenkelBaumann07,HenkelBaumann07b,BaumannEtAl07} the generalised version of LSI makes the following predictions:
\begin{enumerate}
\item The scaling function $\mathcal{R}$ \textit{factorises} into two parts
	\begin{equation}
	\mathcal{R} \left(\frac{t}{s},\, \frac{{\vec r}}{(t-s)^{1/z}}\right) = 
	f_R\left(\frac{t}{s}\right)\,\mathcal{F}\left(\frac{{\vec r}}{(t-s)^{1/z}}\right) \,,
	\end{equation}
	where we may choose $\mathcal{F}(0)=1$.\vspace{5mm}
\item The first factor in this expression, which determines the form of the autoresponse $R(t,s)=s^{-1-a}f_R(t/s)$, again factorises into two power laws of the form
	\begin{equation}
	f_R\left(\frac{t}{s}\right)\;=\;r_0\,\left(\frac{t}{s}\right)^{1+a'-\lambda_R/z}\,
	\left(\frac{t}{s}-1 \right)^{-1-a'}\,,
	\end{equation}
	where $r_0$ is a proportionality constant, $a'$ is another exponent which is generally different from $a$, and $\lambda_R$ is the so-called autoresponse exponent.\vspace{5mm}
\item The second factor, which determines the spatial profile of the response, can be expressed as an integral
	\begin{equation}
	\mathcal{F}^{(\alpha,\beta)}({\vec u})\;=\;
	\int\frac{{\rm d}^d k}{(2\pi)^d}\, |{\vec k}|^\beta \,
	\exp(i{\vec u}\cdot{\vec k}-\alpha|{\vec k}|^z)
	\label{IntegralExpression}
	\end{equation}
	parametrised by a dimensionful non-universal parameter $\alpha$ and another exponent~$\beta$.\vspace{5mm}
\end{enumerate}

\noindent
Baumann {\it et al.} tested these predictions indirectly by measuring the integrated spatio-temporal response and autocorrelation function, finding a convincing agreement. However, this result is surprising because the integral reminds of a L{\'e}vy-stable distribution generated by anomalous diffusion. Such distributions are expected to occur in systems with algebraically distributed long-range interactions \textit{in space}, typically in the range $1<z<2$. The present model, however, does not fall into this category because all interactions are local. It is this disturbing discrepancy which motivated the present work.

In the following we demonstrate that integral expression~(\ref{IntegralExpression}) is \textit{in principle} incom\-patible with the response function of the disordered Ising model in 2D evolving by heat bath dynamics. To see this, one has to evaluate the integral~(\ref{IntegralExpression}) in two dimensions using polar coordinates
\begin{eqnarray}
\mathcal{F}^{(\alpha,\beta)}({\vec u}) \nonumber
&=& \frac{1}{4\pi^2}\int_0^\infty {\rm d}k \, k\,k^\beta
e^{-\alpha k^z}\int_0^\pi{\rm d}\phi \,e^{iuk \cos{\phi}} \\
&=&\frac{1}{4\pi}\int_0^\infty {\rm d}k \, k^{1+\beta}
e^{-\alpha k^z}J_0(uk)\,,
\end{eqnarray}
where $J_0$ denotes the Bessel function of the first kind. The remaining integral can be evaluated exactly for certain rational values of $z$. As an example let us consider the special case $z=4$. Here we obtain the result
\begin{eqnarray}
\mathcal{F}^{(\alpha,\beta)}({\vec u}) &=&\;\;\;\;\;\; \nonumber
\frac{\alpha ^{-\frac{\beta }{4}-\frac{1}{2}} \Gamma \left(\frac{\beta +2}{4}\right) \,
   _1F_3\left(\frac{\beta }{4}+\frac{1}{2};\frac{1}{2},\frac{1}{2},1;\frac{u^4}{256 \alpha }\right)}{16
   \pi }
\\& &\;-\;\frac{u^2 \alpha ^{-\frac{\beta }{4}-1} \Gamma \left(\frac{\beta }{4}+1\right) \,
   _1F_3\left(\frac{\beta }{4}+1;1,\frac{3}{2},\frac{3}{2};\frac{u^4}{256 \alpha }\right)}{64 \pi }\,,
\end{eqnarray}
where $_p F_q\left(\left\{a_1,\ldots,a_p\right\};\left\{b_1,\ldots ,b_q\right\};z\right)$ are generalised hypergeometric functions. Inser\-ting the numerically determined values $\alpha=0.15(2)$ and $\beta=-0.04(10)$ reported in~\cite{BaumannEtAl07}, this function displays an oscillatory behaviour, becoming negative for certain values of~$u$. However, as mentioned before, the response function of the disordered ferromagnetic Ising model evolving by heat bath dynamics is strictly positive. Hence we can conclude that the predictions of LSI, as expressed by Eq.~(\ref{IntegralExpression}), cannot be applied to the disordered Ising model studied here.

%======================================================
\begin{figure}[t]
\begin{center}
\includegraphics[width=100mm]{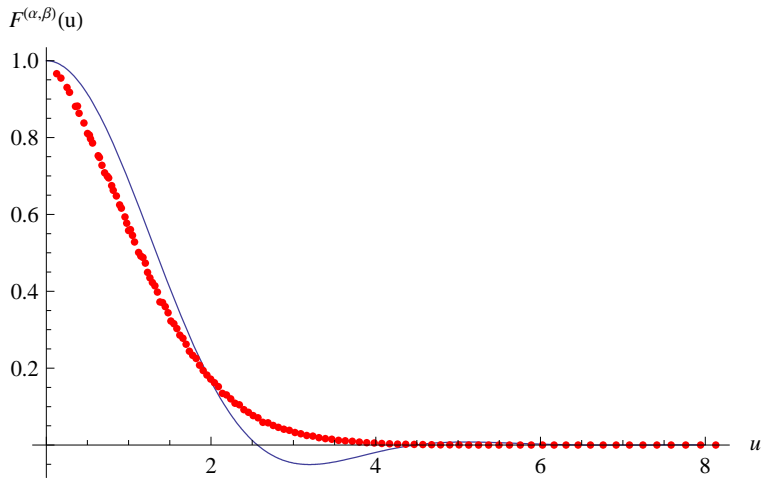}
\end{center}
\vspace{-2mm}
\caption{\label{fig:lsifailure}\small
Prediction of LSI for the scaling function $\mathcal{F}^{(\alpha,\beta)}(u)$ for $\alpha=0.15$ and $\beta=-0.04$ (blue line) compared with numerical data of simulations at temperature $T=1$ below $T_c$ (red dots) for $z=4$. The numerical profile has been obtained by using the algorithm described in Appendix A, with $s=512$, $t=32768$ on a system with $N=750^2$ sites.}
\end{figure}
%======================================================

This incompatibility can be supported by a direct numerical measurement of the response profile. As already observed in the homogeneous kinetic Ising model~\cite{PleimlingGambassi05,CorberiEtAl05,LippielloEtAl06,CorberiEtAl07}, the full response function resolves the deviations from LSI much better than the integrated response (see Appendices for technical details). Fig.~\ref{fig:lsifailure} shows the numerically measured profile and the analytical profile predicted by LSI. As can be seen, the profiles differ significantly.

%==========================================================================
\section{An alternative approach}
%==========================================================================

In the following I suggest an alternative approach which is guided by phenomenological arguments in the spirit of Paul, Puri and Rieger~\cite{PaulEtAl04}. Starting point is the observation that it makes a big difference whether a local perturbation by an external field flips a spin in the interior of an ordered domain or a spin adjacent to a domain wall. In the interior of a domain the generated minority island shrinks extremely fast driven by surface tension so that its contribution to the response function is extremely short-ranged in time and can be neglected. Spin flips at the boundaries, however, may displace a domain wall and thus have a much longer life time. In other words, the response to a local perturbation is always confined to the domain walls and thus propagates in the same way as the domain walls move in space. Without quenched disorder, the domain walls propagate and merge diffusively, hence the local perturbation spreads in space like a random walk, leading to a Gaussian profile.

Quenched disorder changes the dynamics of the domain walls drastically. In particular, domain walls in a disordered environment may be pinned along lines of strongly coupled sites~\cite{PaulEtAl04}. Such a pinned domain wall remains immobile for a certain waiting time. In a scale-free system it is reasonable to assume that these waiting times $\Delta t$ are algebraically distributed as $P(\Delta t)\sim (\Delta t)^{-1-\kappa}$ with $\kappa<1$, leading effectively to a subdiffusive dynamics. With this physical picture in mind it is near at hand to conjecture that the response to a local perturbation in a disordered ferromagnet spreads essentially in the same way as a short-ranged random walk with algebraically distributed waiting times between subsequent random moves. 

To solve this problem, let us first introduce a parameter $\tau$ which is proportional to the number of diffusive moves performed so far. Initially the process starts with $t=\tau=0$. As time proceeds one may ask the question how the elapsed time $t$ is distributed for given $\tau$. This distribution $P(t|\tau)$ evolves by according to the equation
\begin{equation}
\label{DGL}
\Bigl(\partial_t^\kappa+\partial_\tau\Bigr) P(t|\tau)\;=\;0
\end{equation}
with the initial condition $P(t|0)=\delta(t)$, where $\partial_t^\kappa$ denotes a temporal fractional derivative which is known to generate algebraically distributed waiting times (see e.g.~\cite{Fogedby94,Hinrichsen07} and references therein). To solve this equation consider the Fourier transform
\begin{equation}
P(t|\tau) \;=\; \frac{1}{\sqrt{2\pi}}\,
\int_{-\infty}^{\infty} {\rm d} \omega \,e^{i \omega t} \, \tilde{P}(\omega|\tau)
\end{equation}
by which Eq.~(\ref{DGL}) is turned into
\begin{equation}
\Bigl((i\omega)^\kappa+\partial_\tau\Bigr) \tilde{P}(\omega|\tau)\;=\;0
\end{equation}
with the solution $\tilde{P}(\omega|\tau)=e^{-\tau (i \omega)^\kappa}$. By means of the inverse Fourier transform one arrives at
\begin{equation}
P(t|\tau)\;=\;\frac{1}{\sqrt{2\pi}}\,\int_{-\infty}^{\infty} {\rm d} \omega\,e^{i \omega t-\tau (i \omega)^\kappa}	\,,
\end{equation}
where we identify $\kappa=2/z$ by dimensional analysis. This integral can be solved only in certain special cases, e.g.
\begin{equation}
P(t|\tau)\;=\;\left\{
\begin{array}{ll}
%%%%%%%%%%%%%%%%%%%%%%%%%
\delta (t-\tau ) & \mbox{ if } z=2 \\[1mm]
%%%%%%%%%%%%%%%%%%%%%%%%%
\frac{\Gamma \left(\frac{4}{3}\right) \, _1F_1\left(\frac{7}{6};\frac{4}{3};-\frac{4 \tau ^3}{27
   t^2}\right) \tau ^2}{2 \sqrt{3} \pi  t^{7/3}}+\frac{\Gamma \left(\frac{2}{3}\right) \,
   _1F_1\left(\frac{5}{6};\frac{2}{3};-\frac{4 \tau ^3}{27 t^2}\right) \tau }
{2 \sqrt{3} \pi  t^{5/3}}
& \mbox{ if } z=3 \\[1mm]
%%%%%%%%%%%%%%%%%%%%%%%%%
\frac{e^{-\frac{\tau ^2}{4 t}} \tau }{2 \sqrt{\pi } t^{3/2}} & \mbox{ if } z=4
%%%%%%%%%%%%%%%%%%%%%%%%%
\end{array}
\right.
\end{equation}
Note that this solution is already properly normalised by $\int_0^\infty {\rm d}t \, P(t|\tau)=1$ and obeys the required initial condition $P(t|0)=\delta(t)$. 

\begin{figure}[t]
\begin{center}
\includegraphics[width=100mm]{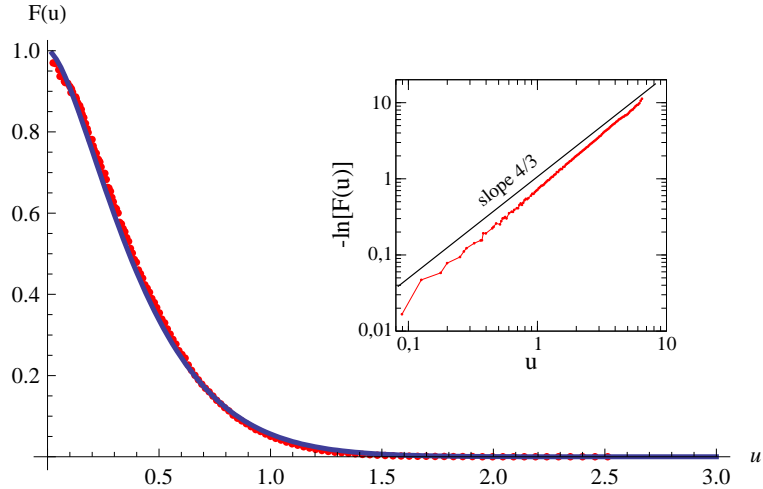}
\end{center}
\caption{\label{fig:compare}\small
Profile according to Eq.~(\ref{Meijer}) (blue) compared to the same numerical data as in Fig.~\ref{fig:lsifailure} (red). The inset illustrates the asymptotic decay, confirming the power $4/3$ predicted in Eq.~(\ref{AsymptoticPower}).}
\end{figure}

The solution gives us the distribution of the physical time $t$ after a given number of diffusive moves $\tau$. However, to compute the response profile one needs to know how many diffusive moves have been performed at a given time $t$. This conditional distribution $P(\tau|t)$ is related to $P(t|\tau)$ by
\begin{equation}
P(\tau|t) \;=\; \frac{{\rm d}}{{\rm d}\tau} \int_0^t {\rm d}t' \, P(t'|\tau)
\end{equation}
giving
\begin{equation}
P(\tau|t)\;=\;\left\{
\begin{array}{ll}
%%%%%%%%%%%%%%%%%%%%%%%%%
\delta (t-\tau ) & \mbox{ if } z=2\\[1mm]
%%%%%%%%%%%%%%%%%%%%%%%%%
\frac{3 t^{2/3} \Gamma \left(\frac{2}{3}\right) \, _1F_1\left(\frac{5}{6};\frac{2}{3};-\frac{4 \tau
   ^3}{27 t^2}\right)+\tau  \Gamma \left(\frac{1}{3}\right) \,
   _1F_1\left(\frac{7}{6};\frac{4}{3};-\frac{4 \tau ^3}{27 t^2}\right)}
{4 \sqrt{3} \pi  t^{4/3}}
& \mbox{ if } z=3 \\[1mm]
%%%%%%%%%%%%%%%%%%%%%%%%%
\frac{e^{-\frac{\tau ^2}{4 t}}}{\sqrt{\pi } \sqrt{t}} & \mbox{ if } z=4
%%%%%%%%%%%%%%%%%%%%%%%%%
\end{array}
\right.
\end{equation}
We now assume that deep in the ageing regime $t\gg s$ the normalised response is just a superposition of Gaussians with different widths according to the varying number of diffusive moves performed so far, i.e.
\begin{equation}
\label{conv}
\frac{R(t,s,{\vec r})}{R(t,s,0)} \;\propto\; \int_0^\infty \, {\rm d}\tau\, e^{-r^2/D\tau} P(\tau|t)\,.
\end{equation}
Here $D$ is the diffusion constant of the random walker which here plays the role of a non-universal metric factor. Unlike the LSI prediction discussed in the previous section, this function is strictly positive. For the special case $z=4$ this integral can be evaluated, giving
\begin{equation}
\label{Meijer}
\frac{R(t,s,{\vec r})}{R(t,s,0)} \;\propto\;  \frac{r^2 G_{0,3}^{3,0}\left(\frac{r^4}{16 t}|
\begin{array}{c}
 -\frac{1}{2},0,0
\end{array}
\right)}{4 \pi  D \sqrt{t}}\,,
\end{equation}
where $G_{p,q}^{m,n}$ denotes the Meijer G-function. This function can be evaluated numerically and may be compared with the normalised numerical data, where $D$ plays the role of a fit parameter. As shown in Fig.~\ref{fig:compare}, both curves are in convincing agreement for $D=2.9$. 

%======================================================
%======================================================

In order to demonstrate that the obtained profile differs from a Gaussian, the integral~(\ref{conv}) may be approximated by the saddle-point method. This yields the expression
\begin{equation}
\frac{R(t,s,{\vec r})}{R(t,s,0)}\;\approx\; \frac{e^{-\frac{3 x^{4/3}}{2^{4/3} D^{2/3} t^{1/3}}} \left(\mbox{erf}\left(\frac{\sqrt{3}
   x^{2/3}}{2^{2/3} D^{1/3} t^{1/6}}\right)+1\right)}{\sqrt{3}}
\end{equation}
which decays far away from the origin as
\begin{equation}
\label{AsymptoticPower}
-\ln \frac{R(t,s,{\vec r})}{R(t,s,0)} \;\sim\; \frac{x^{4/3}}{t^{1/3}}\,,
\end{equation}
confirming non-Gaussian behaviour. As shown in the inset of Fig.~\ref{fig:compare}, the exponent $4/3$ is in fair agreement with the numerical results.

%==========================================================================
\section{Summary}
%==========================================================================

The work by Baumann {\it et al.} was guided by the expectation that LSI generalised to $z\neq 2$ should be a generic theory for ageing phenomena. The disordered Ising model seemed to be a natural candidate to test this expectation in the range $2<z<4$. Studying the integrated response, the authors found an almost perfect coincidence. However, as demonstrated in the present work, a \textit{direct} measurement of the response function reveals significant deviations. Moreover, the predicted response function becomes negative for certain distances while the actual response of the Ising model has to be strictly positive.

In the second part of the paper I demonstrated that another fractional differential equation, which is local in space but non-local in time, gives results in fair agreement with the numerical data. I do not claim that this differential equation has any fundamental meaning, the only purpose is to demonstrate that a simple approach guided by physical intuition in the spirit of Ref.~\cite{PaulEtAl04} leads to a much better agreement with the numerical data. Another possible explanation in terms of super-ageing was proposed by Paul \textit{et al.} in Ref.~\cite{PaulEtAl07} and it would be interesting to investigate how closely these approaches are related. The question whether it is possible to construct a different representation of LSI describing the disordered Ising model remains open.

Finally it should be mentioned the present findings do not support a recently suggested hypothesis of super-universality~\cite{SiciliaEtAl07}, stating that domain growth in homogeneous and weakly disordered Ising systems is described by identical scaling forms.

\noindent
\textbf{Acknowledgements:}\\
I would like to thank M. Henkel for interesting discussions and useful comments. I also thank the referees for pointing out related work.
\vspace{7mm}

%==========================================================================
\appendix
\section{Detailed description of the simulation algorithm}
%==========================================================================

Previous studies on the homogeneous Ising model have shown that the integrated response is not a suitable quantity to detect deviations from LSI, instead one has to measure the full response function. A direct measurement of the response function is notoriously difficult to perform because of strong fluctuations. As a way out, Chatelain proposed to relate the response function to a certain correlation function measured in a system without external field~\cite{Chatelain03}. Based on this idea several authors devised powerful algorithms to measure the response function in the Ising model~\cite{RicciTersenghi03,LippielloEtAl06,CorberiEtAl07}.

Here I present an alternative algorithm for the Ising model with standard heat bath dynamics, where perturbations at \textit{all} sites are processed simultaneously -- a trick which allows one to obtain reasonable statistical averages within short time. A similar algorithm was already used for the contact process (see Ref.~\cite{Hinrichsen06}). In the present case the algorithm is somewhat more complicated, as will be explained in the following. However, the algorithm is conceptually simpler as previous approaches in so far as it measures the response function directly without relying on the non-trivial proof by Chatelain. 

The standard kinetic Ising model with heat bath dynamics is defined as follows. Each site $i$ of a $d$-dimensional lattice is associated with classical variables $\sigma_i = \pm 1$ which represents the orientation of a local spin. Initially the system is prepared in a disordered state, where all spins independently take the values $\pm 1$ with equal probability. Then, starting at time $t=0$, the system evolves at finite temperature $\beta=1/k_BT$ by random-sequential updates according to the following dynamical rules:
\begin{itemize}
\item Select a random site $i$.
\item Calculate the local field $h_i=\sum_j\sigma_j$, where $j$ runs over the $2d$ nearest neighbours of site $i$,      
      respecting the chosen boundary conditions.
\item Generate a random number $z\in (0,1)$ and set
	\begin{equation}
	\sigma_i^{\rm (new)} \;:=\; \left\{ \begin{array}{ll} +1&\mbox{ if } z<\frac12\bigl(1+\tanh(\beta h_i)\bigr)\\
	-1 & \mbox{ otherwise } \end{array}\right.\,.
	\end{equation}
\end{itemize}
As usual in models with random sequential updates, the time variable $t$ is incremented by $1/N$, where $N=L^d$ denotes the total number of sites. 

\noindent
Obviously, the third step can be reformulated equivalently as follows:
\begin{itemize}
\item Define a threshold field 
	\begin{equation}
	\label{threshold}
	h_i^{(c)}:=\beta^{-1}\mbox{atanh}(2z-1)\,,
	\end{equation}
	where $z\in (0,1)$ is a random number, and set
	\begin{equation}
	\sigma_i^{(new)} \;:=\; \left\{ \begin{array}{ll} +1&\mbox{ if } h_i>h_i^{(c)}\\
	-1 & \mbox{ otherwise } \end{array}\right.
	\end{equation}
\end{itemize}
As we will see below, this formulation of the update rule is more convenient for a measurement of the autoresponse function.

In order to measure the autoresponse function $R(t,s)$ directly we would have to apply a weak external field $0<h_i^{\rm (ext)}\ll 1$ at time $s$ locally at a randomly chosen site~$i$ and to study how much the ensemble average of the local magnetisation $\langle \sigma_j(t) \rangle$ at a different site~$j$ changes at some later time $t>s$. However, such a measurement is hard to perform since a localised weak field leads only to occasional spin flips in the future so that the average response to such a \textit{single} spin flip is hard to quantify. This is probably the reason why most researches avoid a direct measurement of the autoresponse function and prefer to study the integrated response. In the kinetic Ising model, however, there are two ways to optimise a direct measurement, namely by (i)~reweighting the probability of spin flips and (ii) simultaneous tagging of all spins in the system.

The algorithm suggested here seems to have a comparable performance as previously applied algorithms without external field based on the paper by Chatelain~\cite{Chatelain03,RicciTersenghi03,LippielloEtAl06,CorberiEtAl07}
Moreover, it relies partially on the same ideas (see below). It would be interesting to compare these algorithms systematically.

\vspace{5mm}

%-----------------------------------------------------------
\noindent{\bf (i) Reweighting spin flip probabilities}\\[3mm]
%-----------------------------------------------------------
%
The first idea is to study the response of an infinitely strong external perturbation instead of a weak one and to extract the limit $h_j^{\rm (ext)}\to 0$ by an appropriate reweighting procedure. Reweighting is a standard technique in Monte-Carlo simulations to enhance the efficiency~(see e.g.~\cite{FerrenbergSwendsen89}). Starting point is the observation that in heat bath dynamics as described above a weak perturbation by a positive localised external field $h_j^{\rm (ext)} \ll 1$ changes the probability to get a positively oriented spin $\sigma_j^{(\rm new)}=+1$ to lowest order as
\begin{eqnarray}
\frac12\bigl(1+\tanh\bigl[\beta (h_j+h_j^{\rm (ext)})\bigr]\bigr) &=& \nonumber
\frac12\bigl(1+\tanh\bigl[\beta h_j\bigr]\bigr) + \Bigl( \frac{\beta}{2} \cosh^{-2}[\beta h_j]\Bigr) h_j^{\rm (ext)}\\
&& + \mathcal{O}([h_j^{\rm (ext)}]^2)\,.
\end{eqnarray}
Therefore, we may equivalently turn up the spin $\sigma_j$ with 100\% probability and compensate this intervention by weighting the resulting response in the ensemble average by the corresponding first-order coefficient. Thus, the reweighting scheme consists of the following steps:
\begin{enumerate}
\item Select random site $i$ at time $s$.
\item Define a weight $w_i=\frac12\beta\cosh^{-2}(\beta h_i)$, where $h_i=\sum_j\sigma_j$ is the local field caused by the nearest neighbours of site $i$ at time $s$, and store them in the memory.
\item Set $\sigma_i:=+1$ by force. 
\item Measure how much this modification increases the magnetisation at site $j$ at some later time $t>s$.
\item Multiply the measured response by the weight factor $w_i$ and then perform the average over many independent realisations.
\end{enumerate}
\vspace{5mm}

%-----------------------------------------------------------
\noindent{\bf (ii) Simultaneous tagging of all possible spin flips caused by an external field}\vspace{3mm}
%-----------------------------------------------------------

\noindent Even with the reweighting procedure described above the measurement of the auto\-response function is still difficult to perform since the response to a perturbation at a \textit{single} site is quite small and fluctuates strongly. However, in the special case of the Ising model with heat bath dynamics it is possible to study the response to a perturbation of \textit{all} spins simultaneously. The reason is that for a given set of random numbers used in the simulation a positive external field can only \textit{increase} the magnetisation in the future, turning spins up but never turning spins down\footnote{As a consequence, the response function $R(t,s,\vec{r})$ is strictly positive, cf. Sect. 2.}. Consequently, the cluster of positively oriented spins generated in a system \textit{without} perturbation by an external field is always a subset of the actual cluster that would have been generated if the perturbation was turned on. Thus, for each site $i$, where the field may have generated a spin flip, it is possible to trace and mark the cluster of all spins in the future that would have been turned up by this perturbation.

\begin{figure*}[b]
\begin{flushright}
\includegraphics[width=131mm]{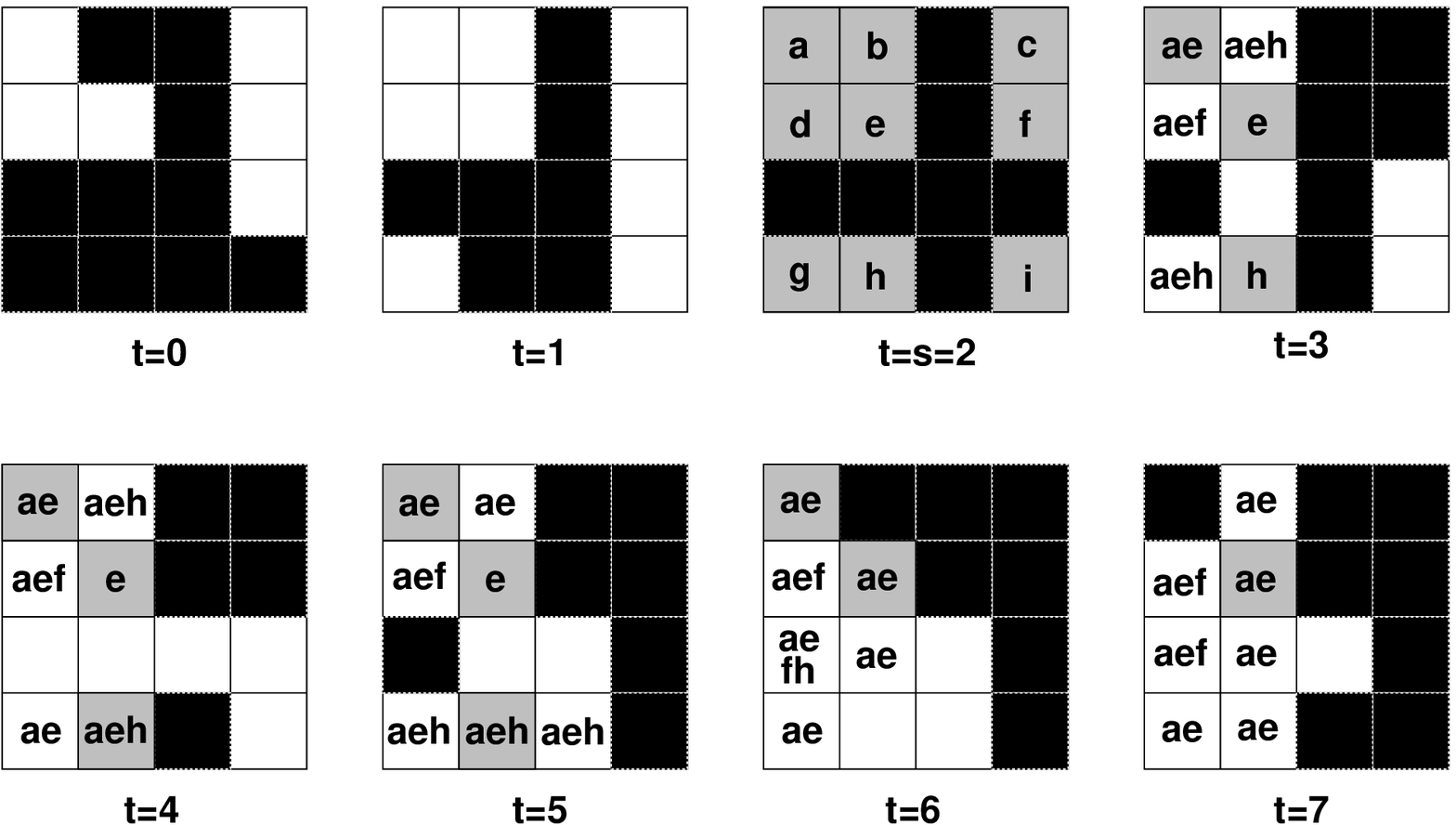}
\end{flushright}
\caption{\label{fig:demo}
Illustration of the algorithm for a small system with $4 \times 4$ sites and periodic boundary conditions without quenched disorder. The figure shows eight snapshots at $t=0,1,\ldots,7$. Initially the system is prepared in a random state. Throughout the whole simulation the system evolves as an ordinary unperturbed Ising model with heat bath dynamics at $T=T_c$, where black boxes represent positive spin variables $\sigma_i=1$ while white or grey boxes represent negative spin variables. At time $t=s=2$ all negatively oriented sites (white boxes) are tagged by individual labels $a,b,\ldots,i$ and the corresponding weights are recorded (here $w_a \simeq 0.024482$, $w_b=w_c=w_d=w_g\simeq 0.110172$, and $w_e=w_f=w_h=w_i\simeq 0.220343)$. Subsequently the simulation tracks how the application of an external local field $h>0$ would have changed the configuration. For example, if an external field would have turned up the spin in the left upper corner at time $t=s=2$ (marked by label $a$) it would have changed the configuration at $t=8$ in such a way that all sites marked by label $a$ (in addition to those marked by black boxes) are positive as well. Note that each site may carry several labels, requiring a dynamically generated set structure in the code. At any time $t>s$ the autoresponse function is the sum over all sites which carry their original label assigned at $t=s$ (indicated by grey boxes), multiplied by the corresponding weight factor. In the present example the autoresponse at $t=7$ is proportional to $w_e$. The measured autoresponse has to be averaged over many independent runs.}
\end{figure*}

The other special property of the Ising model with heat bath dynamics, which is exploited by the algorithm, is the circumstance that these clusters do not interact. More specifically, if the external field had generated two spin flips, the resulting cluster would just be the union of the respective individual clusters. This allows one to tag all spins at time $s$ by non-interfering labels and to identify the percolation cluster for each of the labels (see Fig.~\ref{fig:demo}). The response of a particular site $j$ is then proportional to the number of its labels (re-weighted as described above). This trick accelerates simulations by a factor of the order $N$,  where $N$ is the total number of sites.\footnote{The trick would also allow one to study $n$-point correlation functions efficiently.}

%---------------------------------------------------------------------
%---------------------------------------------------------------------

The tagging algorithm can be implemented as follows. In addition to the local spin variables $\sigma_i$ we attach to each site $i$ a dynamically generated set $S_i$ of labels. For example, in C++ such a dynamical set is provided by the class {\tt set<...> S[N]} of the standard template library STL. Initially all these sets are empty ($S_i=\emptyset$) and the simulation evolves as usual. At time $s$, however, each site $i$ is tagged by an individual label $\Lambda_i$ and the weight $w_i$ is recorded as described above. Therefore, at time $s$ each set $S_i$ has exactly one element, namely, $S_i=\{\Lambda_i\}$. During the subsequent simulation the spin variables $\sigma_i$ evolve according to the usual heat bath dynamics as if there was no perturbation. However, in each update step the following additional steps are carried out
\begin{itemize}
\item When updating spin $i$, first clear the corresponding set by setting $S_i=\emptyset$ and let $U_i=\bigcup_{j}S_j$ be the union of all tags carried by the nearest neighbours. \vspace{2mm}

\item For all labels $\Lambda \in U_i$ count the number $n_{\scriptstyle \Lambda}$ of nearest neighbours which are oriented in negative direction $\sigma_j=-1$ \textit{and} carry the tag $\Lambda$. Obviously, if an external field had flipped the spin corresponding to the label $\Lambda$ at time $s$ it would have increased the local field $h_j$ in the present update by $2n_{\scriptstyle \Lambda}$.\vspace{2mm}

\item If $h_j<h_j^{(c)}$ and  $h_j+2n_{\scriptstyle \Lambda}>h_j^{(c)}$ add label $\Lambda$ to set $S_i$.\vspace{2mm}
\end{itemize}
The autoresponse function $R(t,s)$ is then proportional to the weighted sum over all sets tagged by their original label, averaged over many independent realisations:
\begin{equation}
R(t,s) \propto \left\langle \frac1N \sum_j w_j \delta_{\Lambda_j \in S_j} \right\rangle
\end{equation}
%
% Although the length of the sets varies dynamically, most of them contain a few elements only so that the simulation time scales with the system size as usual.
% \vspace{5mm}

%-----------------------------------------------------------
\section{Excerpt of the C++ code}
%-----------------------------------------------------------
%
\noindent
The following source-code fragment illustrates how the algorithm outlined above can be implemented on a computer. The 2d lattice is stored linearly (row by row) with spins {\tt s[i]} and indices $i$ running from $0$ to $N-1$, where $N=L^2$ is the number of sites. The code requires the existence of a few helper functions which are not listed here, namely:
\begin{itemize}
\item  {\tt InitLattice()} initialises the lattice in a disordered state.
\item {\tt NN(i,d)} with $d=0,1,2,3$ computes the index of the four nearest neighbours of site $i$. 
\item  {\tt TruncatedDistance(i,j)} computes the geometrical distance between sites $i$ and $j$ rounded to an integer.
\item {\tt h(i)} computes the local field at site $i$, where the disordered couplings enter.
\end{itemize}

The following code fragment performs a single run and computes the response {\tt R[r]} at distance $r$ measured at time $t_2$ caused by a perturbation at time $t_1$.

\begin{footnotesize}\begin{verbatim}
#include <set>
#include <map>

const int L=750;                        // system size
const int N=L*L;                        // number of sites
const int t1=512;                       // time of perturbation
const int t2=S*64;                      // time at which the response is recorded
const int T=1;                          // temperature (k_B=1)

...

int s[N];                               // the lattice ss s[i]=+1,-1
double w[N];                            // weight for each perturbed site
double R[L];                            // the response profile measured at time T

set<int> label[N];                      // set of labels at each site
set<int>::iterator p;                   // iterator over all labels
...

InitLattice();                          // Initialize the lattice
for (int t=0; t<=t2; t++) {             // time loop from t=0 to t=t2
        if (t==t1) for (int n=0; n<N; ++n) {             // at t==t1:
                if (s[n]==-1) label[n].insert(n);        // assign labels
                w[n]=0.5/T*pow(cosh(h(n)/T),-2.0);       // assign weights
                }
        for (int v=0; v<N; ++v) {       // perform N random seq. updates
                int i=rndint(N);        // select random site  0...N-1
                label[i].clear();       // delete its labels
                double z=rnd();         // draw random number in [0,1]
                if (2*z<1+tanh(h(i)/T)){// if the field is strong enough
                        s[n]=1;         // turn up the spin        
                else    {
                        s[n]=-1;        // else turn it down

                        // Compute the magn. field generated by each label
                        // at the neighboring spin-down sites:

                        map<int,double> labelfield;
                        map<int,double>::iterator k;

                        // Loop over all neighboring down spins:
                        for (int dir=0; dir<4; dir++) if (S(i,dir)==-1) {
                                int j = NN(n,dir);
                                for (p=label[j].begin(); p!=label[j].end(); p++)
                                        labelfield[*p]+=2*J(i,dir);
                                }

                        // Add all labels that would have created a spin flip:
                        for (k=labelfield.begin(); k!=labelfield.end(); ++k) 
                                if (2*z < 1+tanh((h(i)+(*k).second)/T)) 
                                        label[i].insert((*k).first);
        }       }       }
// Compute response profile produced by a single run
for (int r=0; r<L; ++r) R[r]=0;        // clear profile
for (int i=0; i<N; ++i) 			
       for (p=label[i].begin(); p!=label[i].end(); p++)
                R[TruncatedDistance(*p,i)] += w[*p]/N; 
...
 
// Perform the average over several runs

...
\end{verbatim}
\end{footnotesize}
\vspace{5mm}

%==========================================================================
% References
%==========================================================================
\newpage
\noindent{\bf References}

% \bibliographystyle{iopart-num}
% \bibliography{/home/hinrichsen/Dateien/Literatur/master}
\providecommand{\newblock}{}

\end{document}